\documentclass[prd,twocolumn,a4paper,superscriptaddress,floatfix]{revtex4}
\pdfoutput=1
\usepackage{graphicx}
\usepackage[caption=false]{subfig}
\usepackage{float}
\usepackage{bbm}
\usepackage[all]{xy}
\usepackage{amsmath}
\usepackage{amssymb}
\usepackage[T1]{fontenc}
\usepackage[utf8]{inputenc}

\def\be{\begin{equation}}
\def\ee{\end{equation}}
\def\ben{\begin{eqnarray}}
\def\een{\end{eqnarray}}
\def\ba{\begin{array}}
\def\ea{\end{array}}

\newcommand{\bq}{\begin{eqnarray}}
\newcommand{\eq}{\end{eqnarray}}
\newcommand{\bes}{\begin{subequations}}
\newcommand{\ees}{\end{subequations}}

\begin{document}

\allowdisplaybreaks[3]

\def\a{\alpha}
\def\b{\beta}
\def\g{\gamma}\def\G{\Gamma}
\def\d{\delta}\def\D{\Delta}
\def\ep{\epsilon}
\def\et{\eta}
\def\z{\zeta}
\def\t{\theta}\def\T{\Theta}
\def\l{\lambda}\def\L{\Lambda}
\def\m{\mu}
\def\f{\phi}\def\F{\Phi}
\def\n{\nu}
\def\p{\psi}\def\P{\Psi}
\def\r{\rho}
\def\s{\sigma}\def\S{\Sigma}
\def\ta{\tau}
\def\x{\chi}
\def\o{\omega}\def\O{\Omega}
\def\k{\kappa}
\def\pa {\partial}
\def\ov{\over}
\def\br{\\}
\def\ud{\underline}

\newcommand{\half}{{\textstyle\frac{1}{2}}}
\newcommand\lsim{\mathrel{\rlap{\lower4pt\hbox{\hskip1pt$\sim$}}\raise1pt\hbox{$<$}}}
\newcommand\gsim{\mathrel{\rlap{\lower4pt\hbox{\hskip1pt$\sim$}}\raise1pt\hbox{$>$}}}
\newcommand\esim{\mathrel{\rlap{\raise2pt\hbox{\hskip0pt$\sim$}}\lower1pt\hbox{$-$}}}
\newcommand{\stl}[1]{
  \mbox{$
  \hspace{0.1em}
  \stackrel{\rule{0.4pt}{0.275ex}
    \hspace{0.40em} \!\!\!
    \overline{
      \hspace{0.06em}
      \vphantom{\rule{0.4pt}{0.0ex}}
      \hphantom{\mbox{$\displaystyle #1$}}
      \hspace{0.06em}
    } \!\!\!
    \hspace{0.40em}\rule{0.4pt}{0.275ex}
  }{#1}
  \hspace{0.2em}$
  }
}

\title{Nematic liquid crystal dynamics under applied electric fields}
\author{B. F. de Oliveira}
\affiliation{Centro de F\'{\i}sica do Porto, Rua do Campo Alegre 687, 4169-007 Porto, Portugal}
\affiliation{Departamento de F\'{\i}sica da Faculdade de Ci\^encias
da Universidade do Porto, Rua do Campo Alegre 687, 4169-007 Porto, Portugal}
\affiliation{Departamento de F\'{\i}sica, Universidade Federal da Para\'{\i}ba
58051-970 Jo\~ao Pessoa, Para\'{\i}ba, Brasil}
\author{P.P. Avelino}
\affiliation{Centro de F\'{\i}sica do Porto, Rua do Campo Alegre 687, 4169-007 Porto, Portugal}
\affiliation{Departamento de F\'{\i}sica da Faculdade de Ci\^encias
da Universidade do Porto, Rua do Campo Alegre 687, 4169-007 Porto, Portugal}
\affiliation{Departamento de F\'{\i}sica, Universidade Federal da Para\'{\i}ba
58051-970 Jo\~ao Pessoa, Para\'{\i}ba, Brasil}
\author{F. Moraes}
\affiliation{Departamento de F\'{\i}sica, Universidade Federal da Para\'{\i}ba
58051-970 Jo\~ao Pessoa, Para\'{\i}ba, Brasil}
\author{J.C.R.E. Oliveira}
\affiliation{Centro de F\'{\i}sica do Porto, Rua do Campo Alegre 687, 4169-007 Porto, Portugal}
\affiliation{Departamento de Engenharia F\'{\i}sica da Faculdade de Engenharia
da Universidade do Porto, Rua Dr. Roberto Frias, s/n, 4200-465 Porto, Portugal}

\begin{abstract}

In this paper we investigate the dynamics of liquid crystal textures in a two-dimensional nematic under 
applied electric fields, using numerical simulations performed using a publicly available LIquid CRystal 
Algorithm (LICRA) developed by the authors. We consider both positive and negative dielectric anisotropies 
and two different possibilities for the orientation of the electric field (parallel and perpendicular 
to the two-dimensional lattice). We determine the effect of an applied electric field pulse on the evolution of 
the characteristic length scale and other properties of the liquid crystal texture network. In particular, we show 
that different types of defects are produced after the electric field is switched on, depending on the orientation 
of the electric field and the sign of the dielectric anisotropy.

\end{abstract}

\pacs{}
\maketitle

\section{Introduction}

Defects in the nematic liquid crystalline phase may appear when the symmetry of the isotropic
phase is broken in a isotropic-nematic phase transition, or as a topological
response to imposed boundary conditions \cite{deGennes}. Confined nematics is a stimulating
field of recent research \cite{PhysRevLett.101.037802, PhysRevLett.101.147801,
PhysRevE.79.021707, PhysRevE.80.051703} due to the possibility of different technological
applications. The similarity with defect formation in the early universe has also turned
liquid crystals into mini laboratories for testing the formation and dynamics of cosmic
topological defects \cite{Science.251.1336, PhysRevLett.83.5030, PinaAvelino:2006ia, 
PhysRevE.75.061704, Avelino:2008ve}. It has 
also led one of us (FM) and co-workers to explore the optical properties of defects in a 
liquid crystal in a geometric setting very much like their cosmic analogues \cite{2005MPLA...20.2561S, 2006EPJE...20..173S, 2007EL.....8046002C, 2008EPJE...25..425S}.

The complexity of the (dis)organization of nematic molecules in the presence of defects limits
severely what can be analysed using analytical tools. Hence, computer simulations are
an important complement in the study of the evolution of liquid crystalline systems with
defects. Lattice simulations have been used extensively in the study of liquid crystals 
(see, for example,  \cite{PhysRevE.60.6831, PhysRevE.66.051705, IEEE.12.1323, bhattacharjee:026707, 
PhysRevE.79.061703}) since the pioneering work of Lebwohl and Lasher \cite{PhysRevA.6.426}. 
An order parameter is usually attached to each point of the lattice with orientational degrees 
of freedom only, so that the liquid does not flow but each molecule is free to orient itself in 
three-dimensional space. The fact that a relatively large number of particles enter the simulation 
is particularly useful to study collective behavior such as phase transitions and systems 
presenting large scale fluctuations and defects.

From the industry point of view, defects may be quite relevant. They are a problem for 
liquid crystal displays since they scatter light, but they are unavoidable in the switching 
of the cells. Consequently, it is interesting, both from scientific and technological 
points of view, to study how applied external electric fields influence the 
formation and dynamics of defects in nematics. External electric or magnetic fields 
couple to the director adding an extra ingredient to an already complex problem 
\cite{PhysRevE.71.061709, PhysRevE.80.031706, JComputTheorNanosci.7.709, 
PhysicaB.405.2118, OptExpress.13.2201,PhysRevLett.101.247801,PhysRevE.81.040701}. 

In this work we simulate the dynamics of defects in a two-dimensional nematic in the 
presence of an applied electric field, after quenching from the isotropic phase. 
The outline of this paper is as follows. In Sec. II we present the Landau-de Genes 
theory which we use to model the relaxational dynamics 
of nematic liquid crystals under applied electric fields. In Sec. III we report on a 
publicly available LIquid CRystal Algorithm (LICRA) developed by the authors and 
describe the specific numerical implementation used in this paper. In Sec. IV the 
results are presented, considering different possibilities for the orientation of the 
electric field and sign of the dielectric anisotropy. We conclude in Sec. V.

\section{Landau-de Gennes theory}

The orientational order of a nematic liquid crystal is described by a symmetric traceless
tensor, ${\bf Q}$, whose components are given by  \cite{deGennes}.
\begin{eqnarray}
\label{qdef}
Q_{\alpha \beta}({\bf r})&=&-\frac13 \delta_{\alpha \beta} + \langle {u_\alpha u_\beta}\rangle
\nonumber \\ &=& -\frac13 \delta_{\alpha \beta} + \int d^2 {\bf u} \, f({\bf r},{\bf u},t)
u_\alpha u_\beta \,,
\end{eqnarray}
where $f({\bf r},{\bf u},t)$ is the the molecular orientational distribution function at the
position ${\bf r}$ and time $t$, measuring the number of molecules with orientation defined by
the unit vector ${\bf u}$ with components $u_\alpha$. The order parameter, ${\bf Q}$, can be
written as a function of the principal values $S$ and $T$ and the eigenvectors ${\bf n}$ (the
director), ${\bf l}$ (the codirector) and ${\bf m}={\bf n} \times {\bf l}$. Its components are
given by
\begin{equation}
Q_{\alpha \beta} = \frac32 S  \stl{n_\alpha n_\beta}- \frac12 T\left(l_\alpha l_\beta -
m_\alpha m_\beta\right)\,,
\end{equation}
where
\begin{equation}
\stl {X_{\alpha \beta}} = \frac12 \left(X_{\alpha \beta}+X_{\beta \alpha}\right)
-\frac13\delta_{\alpha \beta}
X_{\gamma \gamma} \,.
\end{equation}

Defining $\theta$ and $\phi$ such that $\cos \theta= {\bf u} \cdot {\bf n}$, $\sin \theta \cos
\phi= {\bf u} \cdot
{\bf l}$ and $\sin \theta \sin \phi= {\bf u} \cdot {\bf m}$, the principal values $S$ and $T$
can be written as
\begin{eqnarray}
S &=&  \langle \cos^2 \theta \rangle - \frac13\,, \\
T &=& \langle {\sin^{2}\theta \cos2\phi}\rangle\,.
\end{eqnarray}
The inner product is defined by ${\bf a} \cdot {\bf b} = a_\gamma b_\gamma$ where we use the
Einstein notation for
the summation of repeated indices. Also, note that in Eq. (\ref{qdef}) $d^2 {\bf u} = \sin^2
\theta \, d\theta \, d \phi$.

The free energy is given by
\begin{equation}
F=\int d^3 {\bf r} \left( \mathcal F_I +  \mathcal F_{II}+  \mathcal F_{III}\right)\,,
\end{equation}
where $\mathcal F_I$ is obtained from a local expansion in powers of rotationally invariant
powers of the order parameter
\begin{equation}
\mathcal F_I= \frac{A}{2} \,  Q_2  +
\frac{B}{3} \, Q_3+ \frac{C}{4} \, Q_2^2 + \frac{D}{6} \, Q_3^2\,,
\end{equation}
$\mathcal F_{II}$ is a non-local contribution arising from rotationally invariant combinations
of gradients of the order parameter
\begin{equation}
\mathcal F_{II}= \frac{R_1}{2} \, \partial_{\alpha}
Q_{\beta\gamma} \, \partial_{\alpha}Q_{\beta\gamma} +
\frac{R_2}{2} \, \partial_{\alpha}Q_{\alpha\beta} \, \partial_{
\gamma}Q_{\beta\gamma}\,,
\end{equation}
and $\mathcal F_{III}$ accounts for the effect of an external electric field ${\bf E}$
\begin{equation}
\mathcal F_{III}=-\frac{\tau_E}{2} Q_{\alpha \beta} E_\alpha E_\beta\,,
\end{equation}
where $\tau_E=2 \Delta \epsilon$ and $\Delta \epsilon$ is the dielectric anisotropy.

Static equilibrium can only be reached for a minimum value of the free energy ($\delta F/
\delta Q_{\mu \nu} = 0$). However, the evolution of the order parameter $\bf Q$ from a given
set of initial conditions is not fully specified by the free energy functionals, and further
assumptions have to be made on how the minimization process will take place. In the absence of
thermal fluctuations and hydrodynamics flow, the time evolution of the order parameter is given
by
\begin{equation}
{\dot Q}_{\alpha\beta}({\bf r}, t) = - \Gamma_{\alpha
\beta\mu\nu} \frac{\delta F}{\delta Q_{\mu\nu}}\,.
\end{equation}
Here the dot represents derivative with respect to the physical time, $t$, and the tensor
\begin{equation}
\Gamma_{\alpha\beta\mu\nu} = \Gamma\left(\delta_{\alpha\mu}\delta_{\beta\nu} +
\delta_{\alpha\nu}\delta_{\beta\mu} - \frac{2}{3}\delta_{\alpha\beta}
\delta_{\mu\nu}\right)\,,
\end{equation}
satisfies $\Gamma_{\alpha\beta\mu\nu} = \Gamma_{\beta\alpha\mu\nu}=\Gamma_{\mu\nu\alpha\beta}$
and $\Gamma_{\alpha\alpha\mu\nu} =0$ thus ensuring that the order parameter $\bf Q$ remains
symmetric and traceless. In the following we shall assume that the kinetic coefficient,
$\Gamma$, is a constant.

The equation of motion can be written more explicitly as
\begin{eqnarray}
\label{Qequation}
{\dot Q}_{\alpha\beta} &=& - \Gamma [(A +
C Q_2)Q_{\alpha\beta} + (B + D Q_3)
\stl{Q_{\alpha\gamma} Q_{\gamma\beta}} \nonumber \\
&& - R_1 Q_{\alpha\beta,\gamma \gamma} -
R_2\stl{Q_{\beta\gamma,\gamma \alpha}}-\tau_E \stl{E_\alpha E_\beta}]\,,
\end{eqnarray}
where a comma denotes a partial derivative.

\section{Numerical implementation}

We solve Eq. (\ref{Qequation}) using a standard second-order finite difference algorithm for
the spatial derivatives and a second order Runge-Kutta method for the time integration. All the
simulations are performed on a two dimensional $256^2$ grid (in the $xy$ plane) with periodic boundary
conditions. The order parameter ${\bf Q}$ has 5 degrees of freedom associated with $S$, $T$,
$\bf n$, $\bf l$ ($\bf n$ accounts for two degrees of freedom). In the simulations, the initial
conditions for $S$ and $T$ are randomly generated, at every grid point, from uniform
distributions in the intervals $[0,1[$ and $[0,S[$, respectively. The director, ${\bf n}$, is
also randomly generated, at every grid point, using the spherical vector distribuitions
routines in the GNU Scientific Library and the codirector, $\bf l$ , was calculated by
randomly choosing a direction perpendicular to ${\bf n}$.

We start our simulations with the electric field switched off, switching it on at a given time,
$t_{\rm on}=400$, and disconnecting it at a later time, $t_{\rm off}=600$. We consider two different
possibilities for the orientation of the applied electric field, perpendicular or parallel to
the plane of the simulation. In each case we consider situations with either positive
($\Delta\epsilon > 0$) or negative ($\Delta \epsilon < 0$) dielectric anisotropies.
The parameters used in the simulations are $\Delta x=\Delta y =1$, $\Delta t=0.1$ and 
$A= -0.1$, $B= -0.5$, $C=2.67$, $R_{1}= 1.0$, $R_{2}=0.0$, $\Gamma = 1.0$, corresponding to a uniaxial nematic 
phase  \cite{bhattacharjee:026707}). The amplitude of the applied rectangular electric field pulse is taken to be 
$E= 0.025 / {\sqrt {|\Delta \epsilon|}}$.

In order to determine the evolution of the characteristic scale of the network we numerically calculate the
function
\begin{equation}
P({\bf k}, t) = \frac{Q_{\alpha\beta}({\bf k}, t) Q_{\beta\alpha}({\bf -k}, t)} 
                {\int{d^3{\bf k} \; Q_{\alpha\beta}({\bf k}, t) Q_{\beta\alpha}({\bf -k}, t)}}\,,
\end{equation}
after setting the infinite wavelength mode $Q_{\alpha\beta}({\bf 0})$ to zero. Here
$Q_{\alpha\beta}({\bf k})$ is the Fourier transform of $Q_{\alpha\beta}({\bf r})$ (${\bf k}$ is
the wavenumber). The characteristic scale, $L$, can then be defined as
\begin{equation}
\frac{1}{L^2}=\langle k^2\rangle = \sum_{{\bf k}} k^2 P({\bf k},t)/\sum_{\bf k}P({\bf k}, t)\,.
\end{equation}

The set of C codes used to perform the simulation are available at {\bf http://faraday.fc.up.pt/licra}
along with Matlab/Octave routines that are used to generate the results visualization. This
software is free and we have named it LICRA (LIquid CRystal Algorithm). LICRA can be 
used to perform numerical simulation of the dynamics of liquid crystal texture networks 
in both two- and three-dimensional nematics. LICRA also accounts for the application of 
an external electric field on the system. Simulation parameters may be modified in the file licra.h 
or directly on the code which is properly commented for this effect. The GNU Compiler 
Collection (gcc) and both  the Fastest Fourier Transform in the West (FFTW) and the GNU 
Scientific Library (GSL) are required in order to compile the source code.

\begin{figure}
 	\subfloat[]{\label{before_cp}\includegraphics[scale= 0.4]{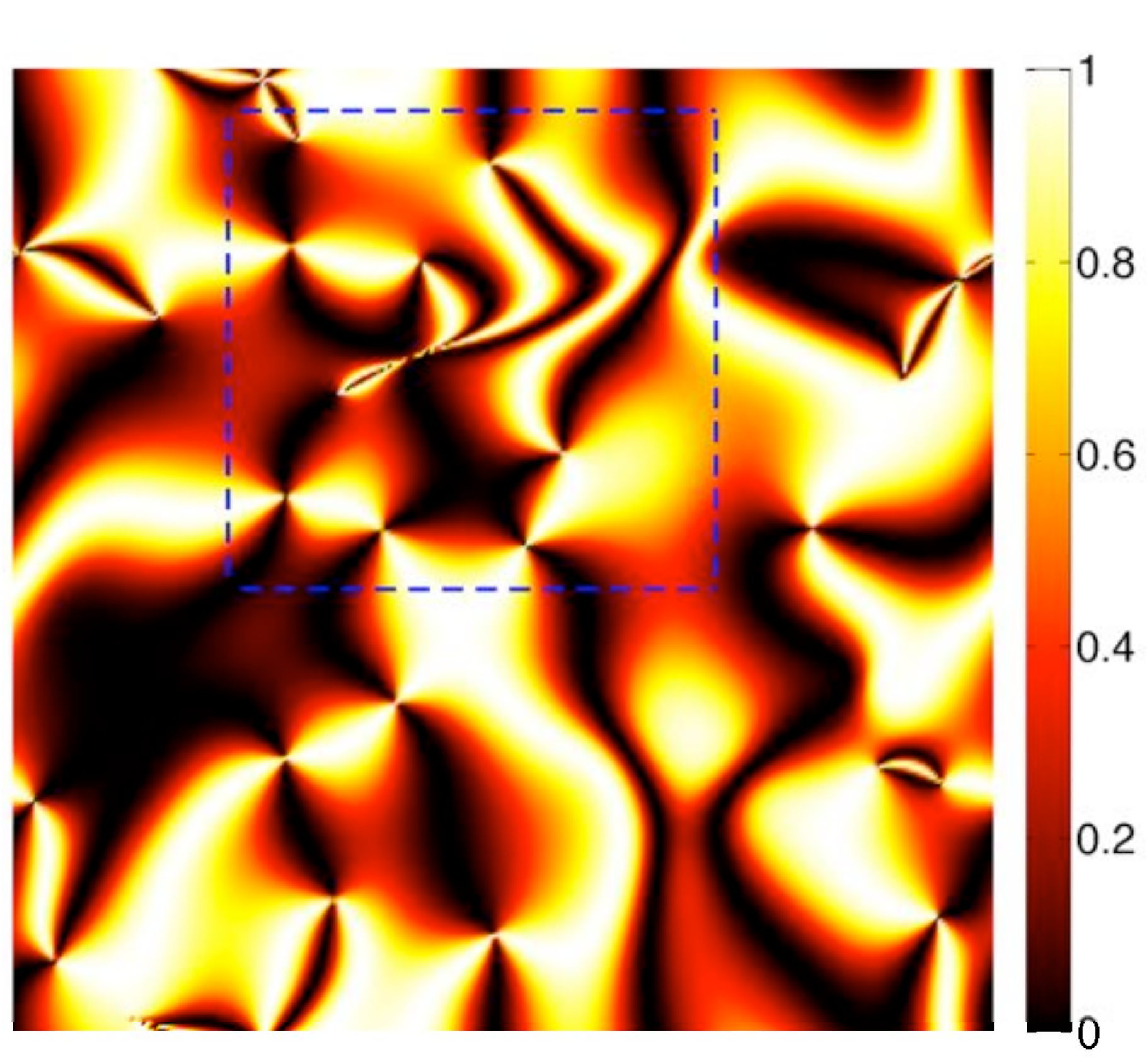}}\\
  \subfloat[]{\label{before_s}\includegraphics[scale= 0.4]{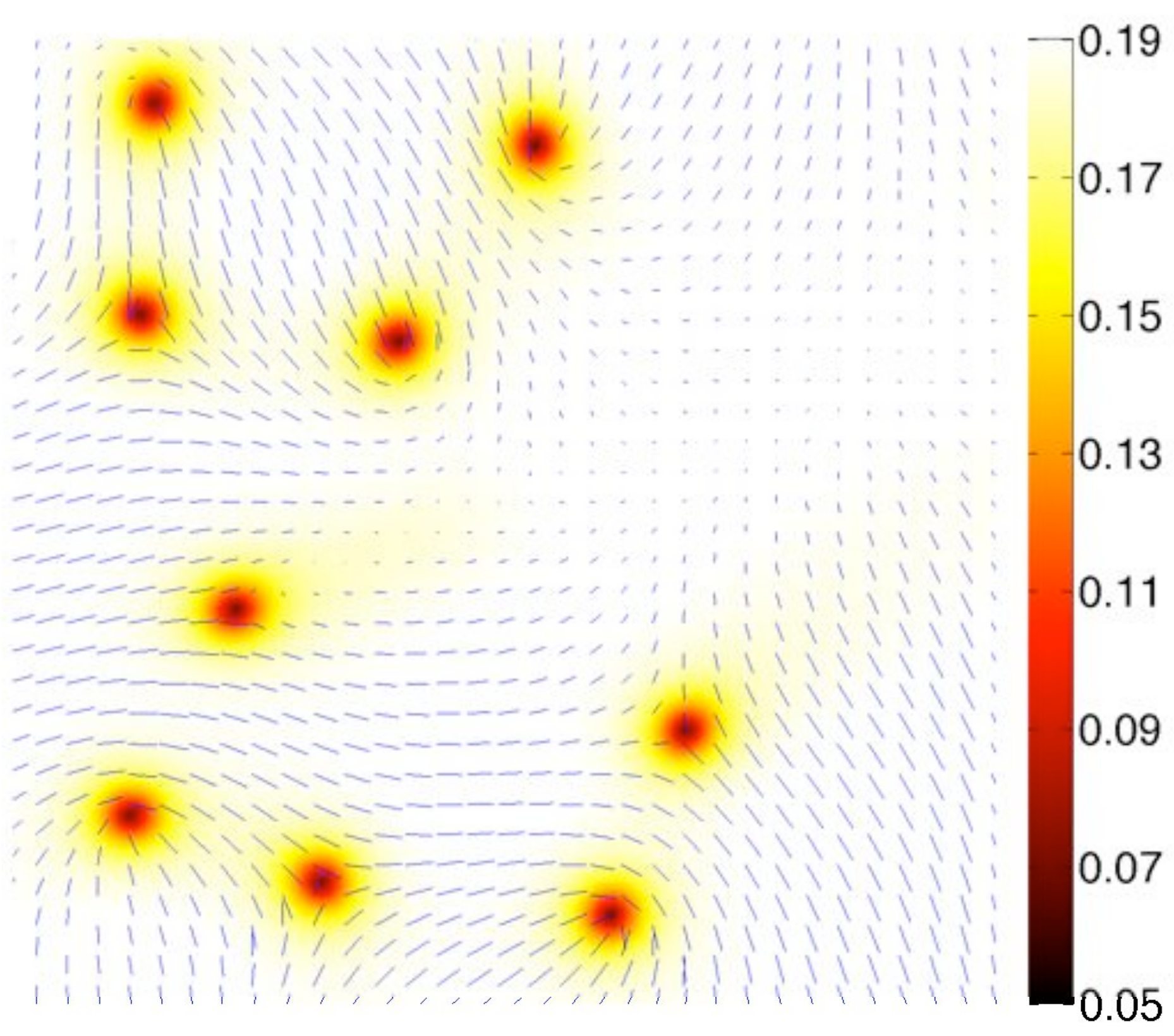}}\\
	\subfloat[]{\label{before_t}\includegraphics[scale= 0.4]{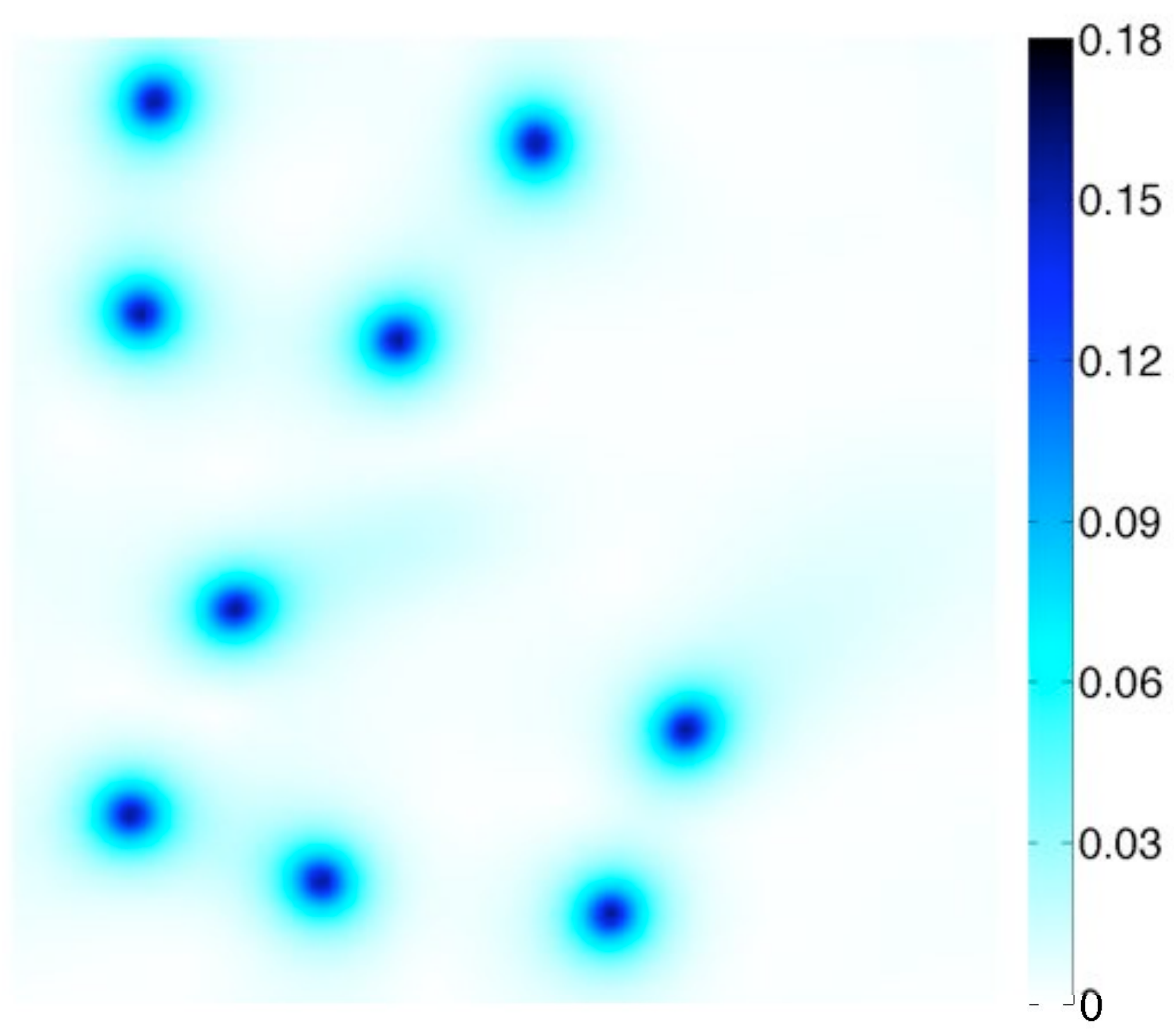}}
\caption{Topological defects in a coarsening nematic liquid crystal before the application of
an external electric field. Fig. \ref{before_cp} represents the value of $\sin^2 (2
\chi)$ at every grid point (note that both
half-integer and integer disclinations appear in the simulations). Fig. \ref{before_s} 
shows the value of the parameter $S$ at every grid point inside the
dashed region, as well as the projection  of the director ${\bf n}$ onto the two dimensional
grid. In Fig. \ref{before_t} the value of the parameter $T$ in same region is given.}
  \label{before}
\end{figure}

\begin{figure}
  \includegraphics[scale= 0.4]{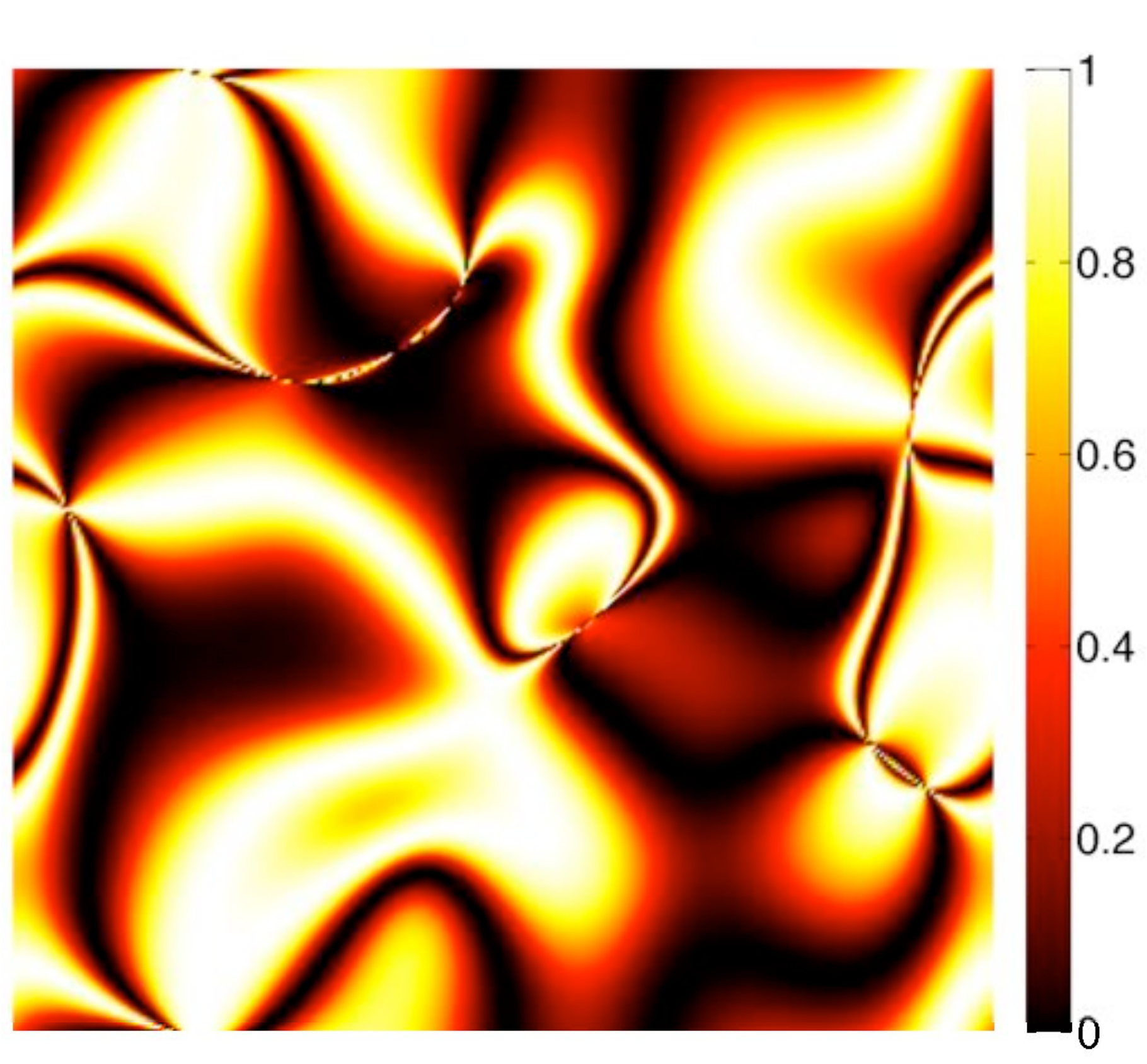}
  \caption{Similar to Fig. \ref{before_cp} but now after the application of an electric field
pulse perpendicular to the lattice plane. The figure shows that, in the case of a positive
dielectric anisotropy ($\Delta \epsilon > 0$), only integer ($\pm 1$) disclinations remain in
the simulation, after the application of the external electric field.}
  \label{pos_per_cp}
\end{figure}

The code outputs contain the results for the order parameter $S$ and $T$, the director and
codirector components at each grid point and the angle between the projection of the director
onto the $x-y$ plane and the $x$ axis. The visualization routines generate the intensity of the
transmitted light after passing thought a nematic liquid crystal sample with crossed
polarizers, the parameters $S$ and $T$, and the evolution of characteristic length scale 
with time.

\section{Results}

Fig. 1 represents a texture network in a coarsening nematic liquid crystal just before the application of
an external electric field at $t = t_{\rm on}$. The value of $\sin^{2}(2 \chi)$ is plotted in Fig. \ref{before_cp}, 
where $\chi$ is the angle of an arbitrary axis on the plane of the simulation with the projection of the director ${\bf
n}$ onto the same plane. Topological defects of both half-integer ($\pm 1/2$) and integer
($\pm 1$) charge appear in the simulations, corresponding to the meeting points of two and four
dark brushes, respectively. Schlieren patterns similar to the ones shown in Fig. (\ref{before_cp}) 
may be obtained experimentally using a linear polarized light microscope with a liquid crystal 
sample placed between crossed polarizers. 

Fig. \ref{before_s} shows the value of $S$ at every grid point inside the dashed region as
well as the projection, onto the two dimensional grid, of the director ${\bf n}$. In Fig. \ref{before_t} 
the value of the parameter $T$ is also shown for the same region. There is a
rapid variation of the parameters $S$ and $T$ at the half-integer charge defect cores while no
significant variation occurs in the case of integer charge defects. The non-zero values of $T$
are due to the biaxiality inside half-integer defect cores \cite{PhysRevLett.59.2582}. It is
also clear that integer charge defects are associated with regions where the projection of the
director ${\bf n}$, onto the two-dimensional grid, is very small.

\begin{figure}
  \subfloat[]{\label{neg_per_cp}\includegraphics[scale= 0.4]{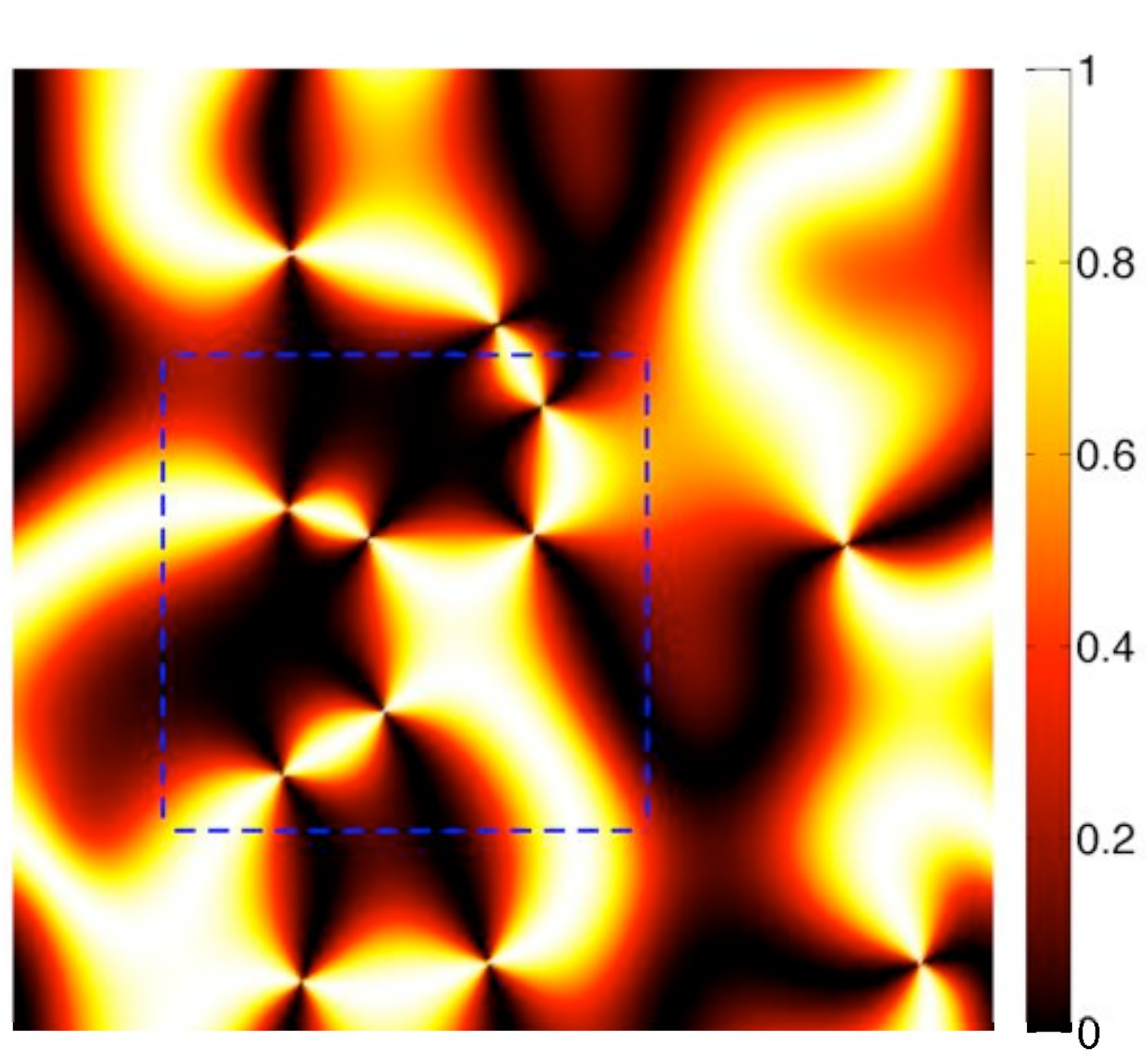}}
        \\
  \subfloat[]{\label{neg_per_s}\includegraphics[scale= 0.4]{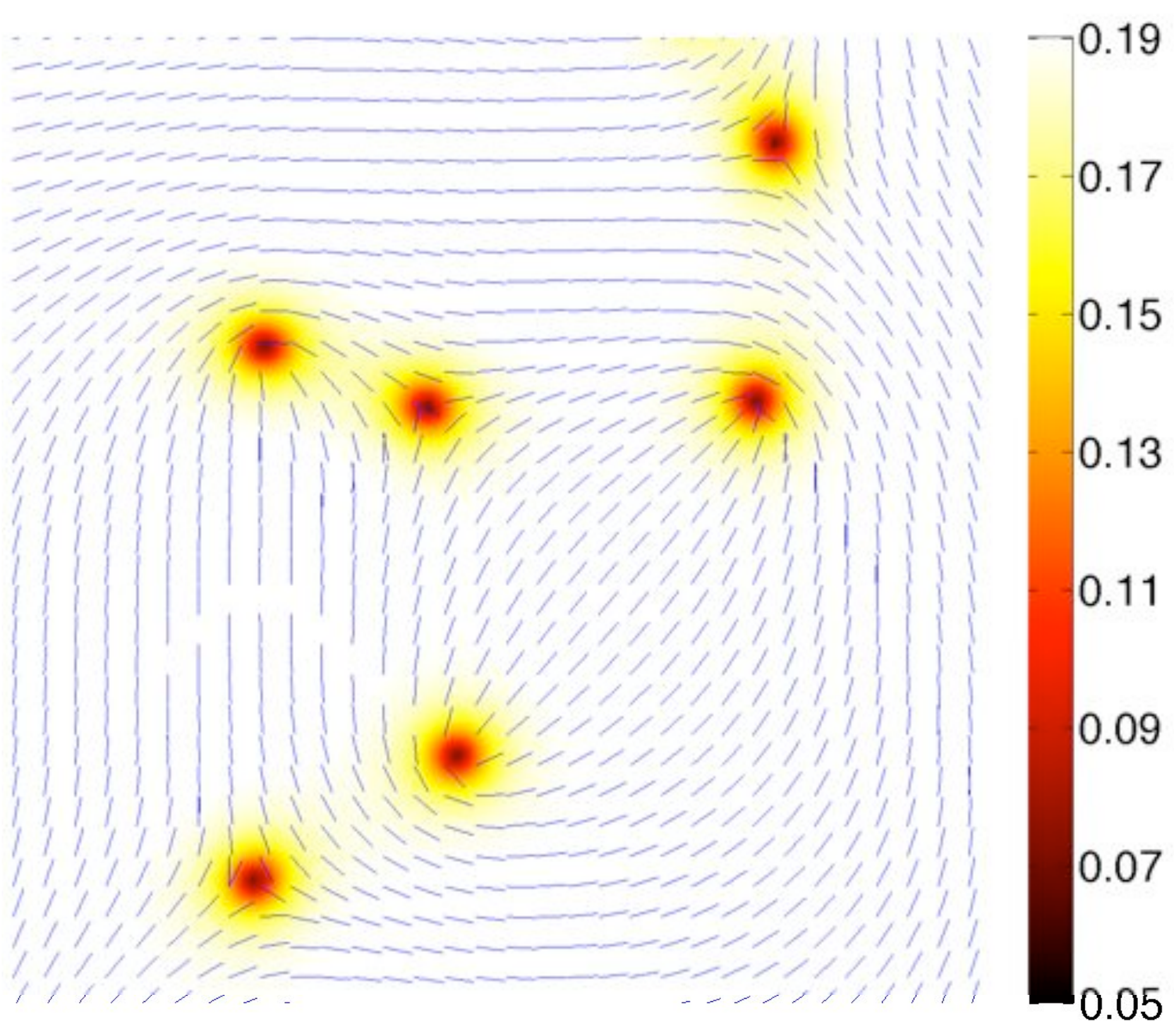}}
        \\
  \caption{Similar to Figs. \ref{before_cp} and \ref{before_s} except that now a negative dielectric
anisotropy ($\Delta \epsilon < 0$) was considered (again with an electric field pulse 
perpendicular to the lattice plane). In this case, only half-integer ($\pm 1/2$)
disclinations survive, after the application of the external electric field.}
  \label{pos_per_cp }
  \label{neg_per}
\end{figure}

\begin{figure}
\includegraphics[scale= 0.4]{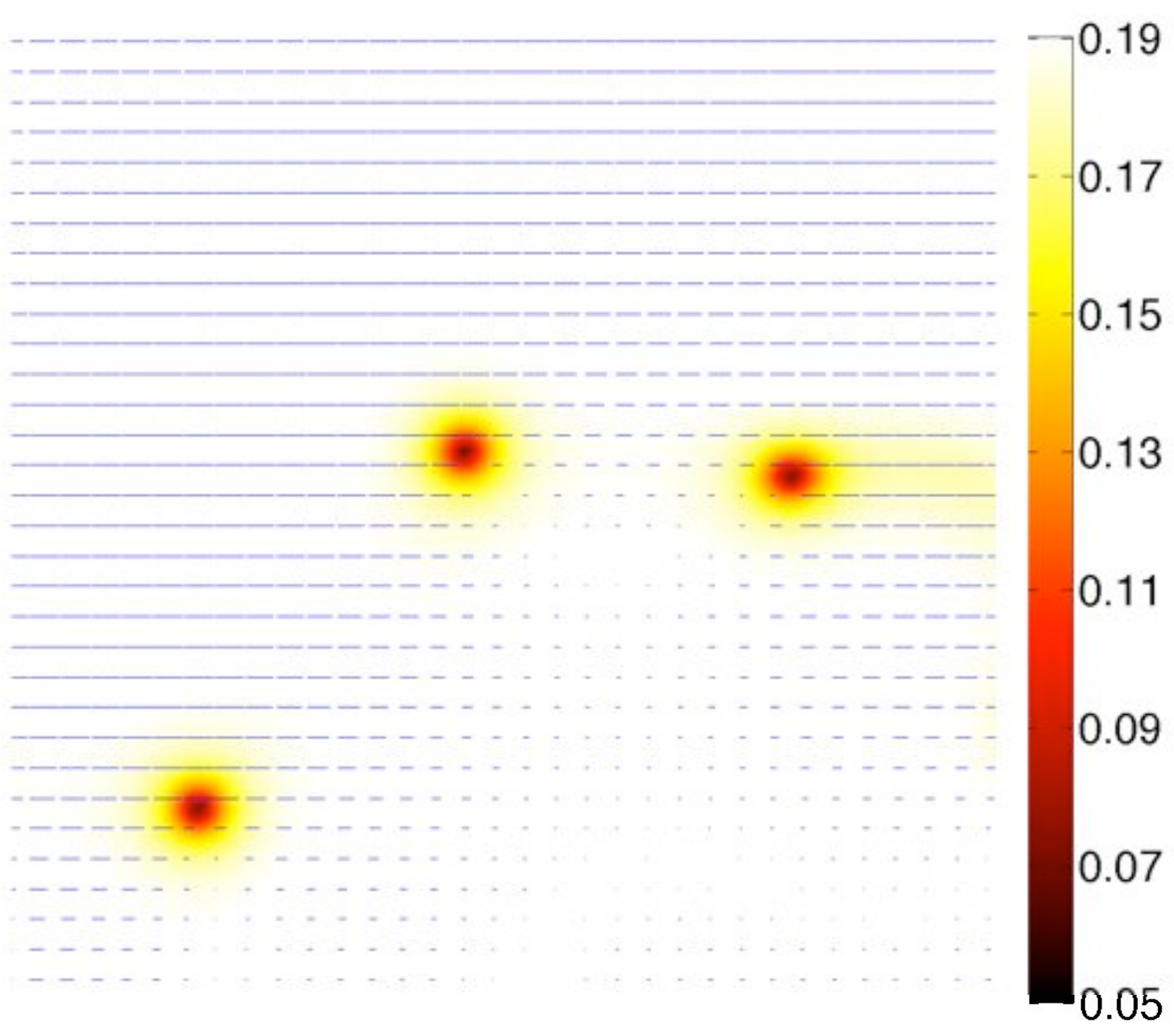} 
\caption{The figure represents the order parameter $S$ as well as the projection of the
director, ${\bf n}$, onto the two-dimensional grid, after the application of an electric field
pulse parallel to the lattice plane. A negative dielectric anisotropy ($\Delta \epsilon <  0$)
was considered in this case. All the defects appearing in the simulations correspond to
half-integer twisted disclinations.}
  \label{neg_par_s}
\end{figure}

\begin{figure}
  \includegraphics[scale= 0.4]{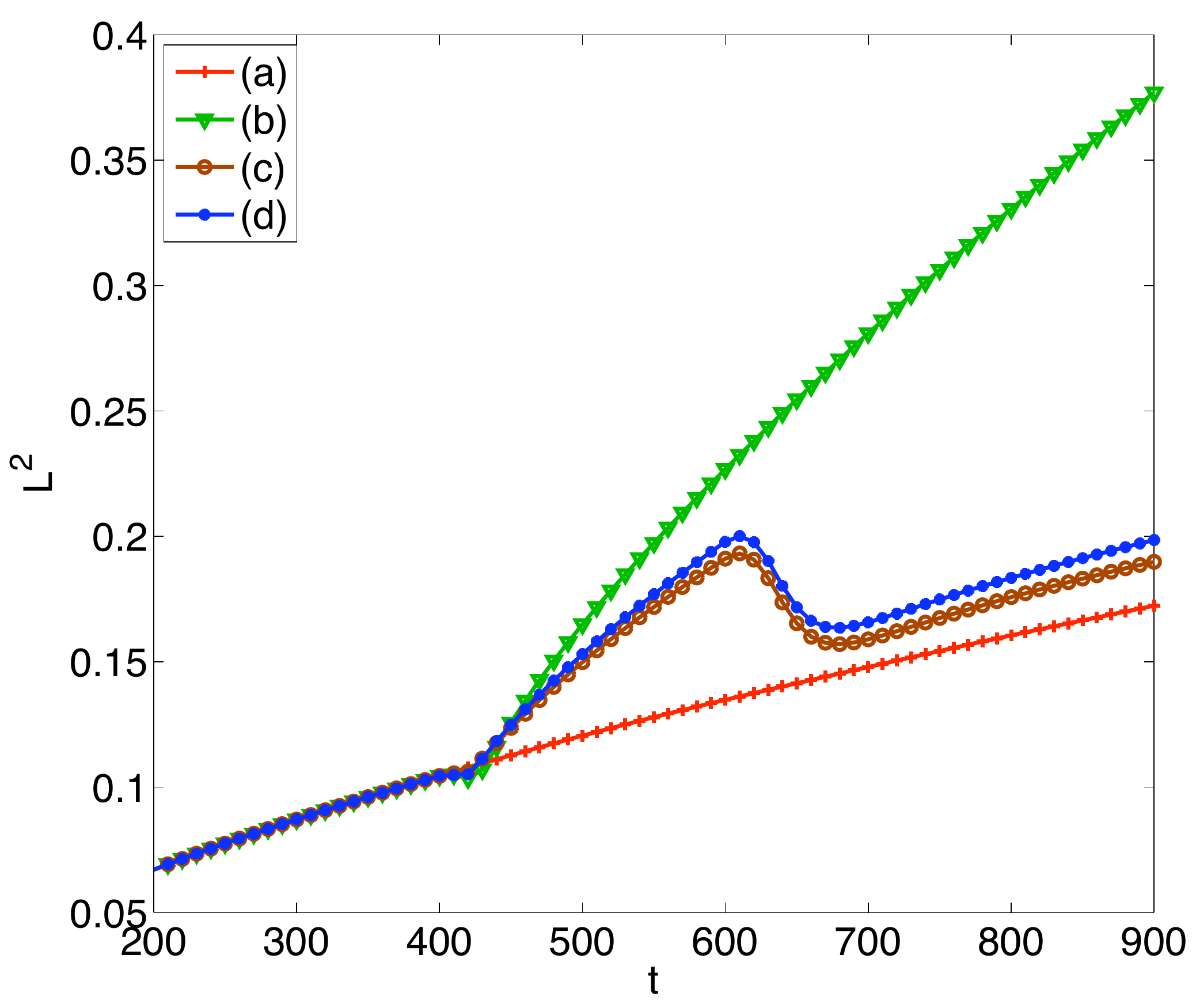}
  \caption{The evolution of the correlation length, $L(t)$, in four different situations: $(a)$
without external electric field, $(b)$ with an electric field perpendicular to the lattice plane 
and $\Delta \epsilon > 0$, $(c)$ the same as $(b)$ but with $\Delta \epsilon < 0$ and $(d)$
with an electric field parallel to the lattice plane and $\Delta \epsilon < 0$. The electric
field is connected when $t=t_{\rm on}=400$ and disconnected when $t=t_{\rm off}=600$. 
}
 \label{lt}
\end{figure}

Figs. 2, 3 and 4 represent a texture network in a coarsening nematic liquid crystal after the 
application of an external electric field ($t = 800 \ (> t_{\rm off})$). Fig. \ref{pos_per_cp} shows the 
value of $\sin^2 (2 \chi)$ at every grid point after the application of an electric field pulse 
perpendicular to the lattice in the case of a positive dielectric anisotropy ($\Delta \epsilon > 0$). 
After the field is switched on, the molecules (and the director ${\bf n}$ at each grid point) tend to 
become aligned parallel to the field, which results in the rapid suppression of half-integer 
disclinations (only integer ones remain in the simulations).
Accordingly, $S$ is roughly constant and $T$ is very close to zero at all network points. In
this case the projection of the director ${\bf n}$, onto the two-dimensional grid, is very
small at every grid point. Apart from the modification of the background values of $S$ and $T$
there is no significant modification to the network evolution when the electric field is
switched off (this can be confirmed in Fig. {\ref{lt}}).

\begin{figure}
	\subfloat[]{\label{pos}\includegraphics[scale= 0.4]{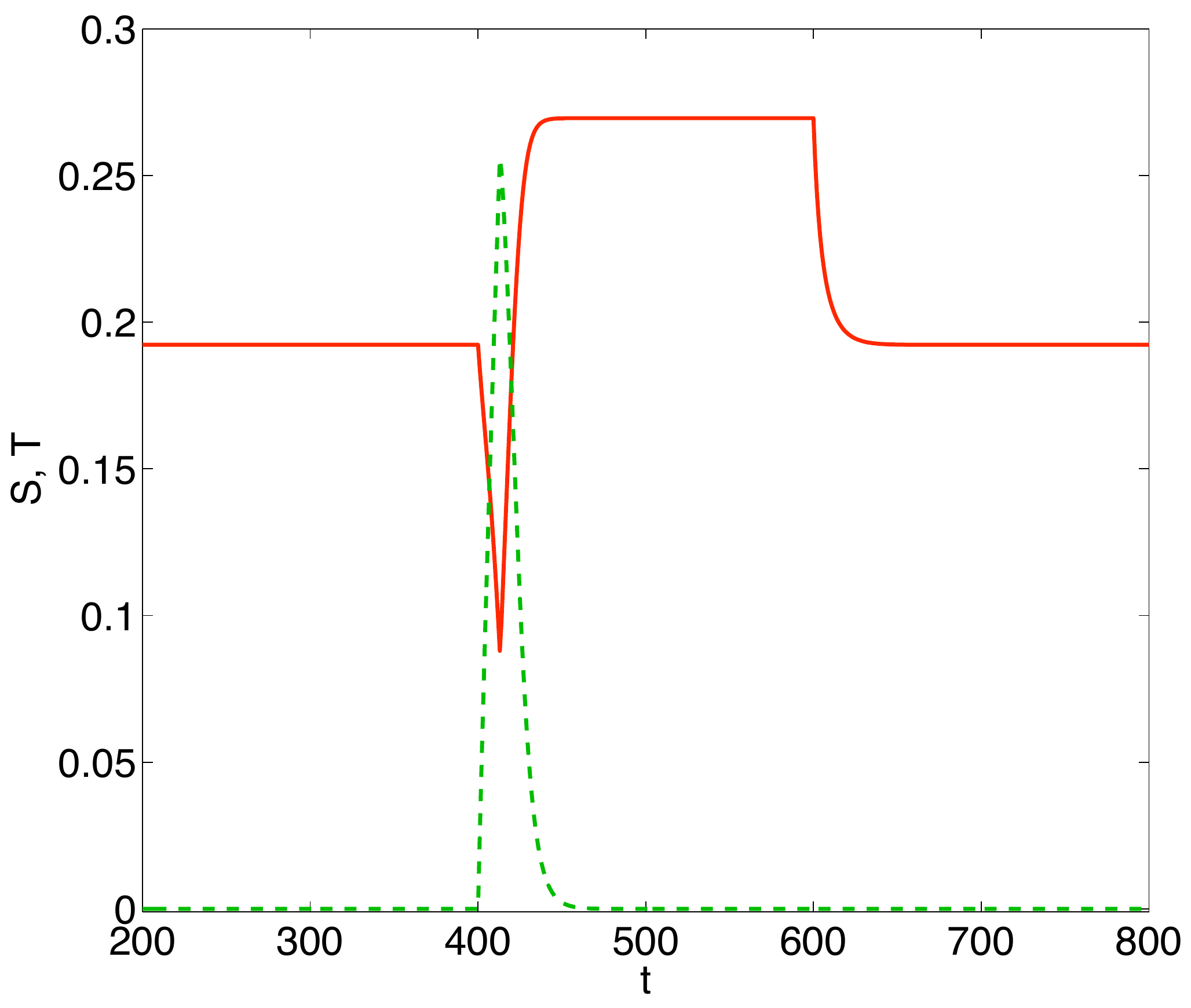}}
	\\
  \subfloat[]{\label{s_and_t_neg}\includegraphics[scale= 0.4]{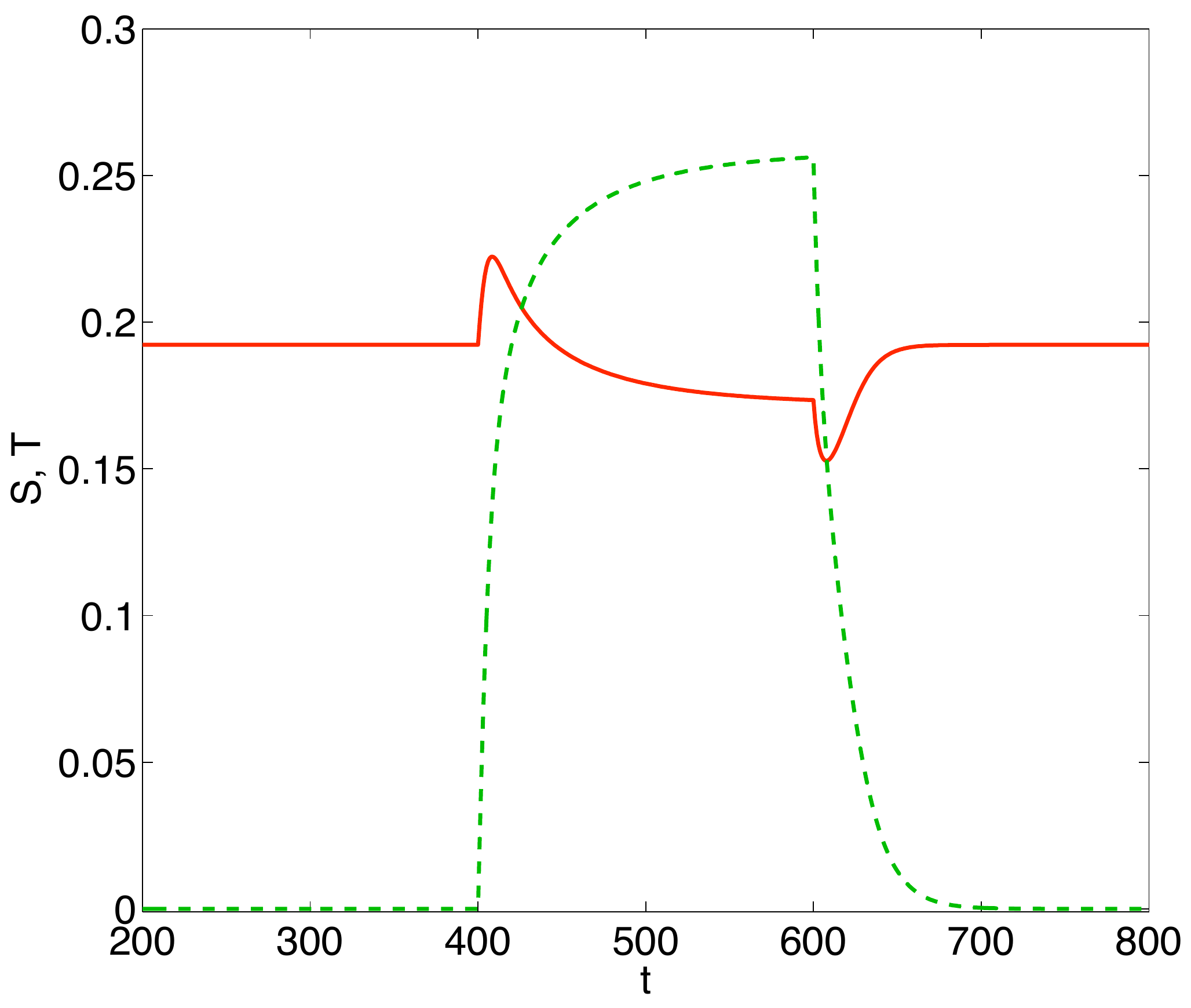}}
	\\	
  \caption{Figs. \ref{pos} and \ref{s_and_t_neg} represent the evolution of the parameters
$S$ and $T$ for the case of an homogeneous director, ${\bf n}$, initially aligned parallel to
the lattice plane considering, respectively, either a positive  ($\Delta \epsilon > 0$) or
negative  ($\Delta \epsilon < 0$) dielectric anisotropy. In both cases an electric field
perpendicular to the plane is connected for $t=t_{\rm on}=400$ and disconnected for $t=t_{\rm off}=600$.}
\end{figure}

\begin{figure}
	\subfloat[]{\label{max}\includegraphics[scale= 0.4]{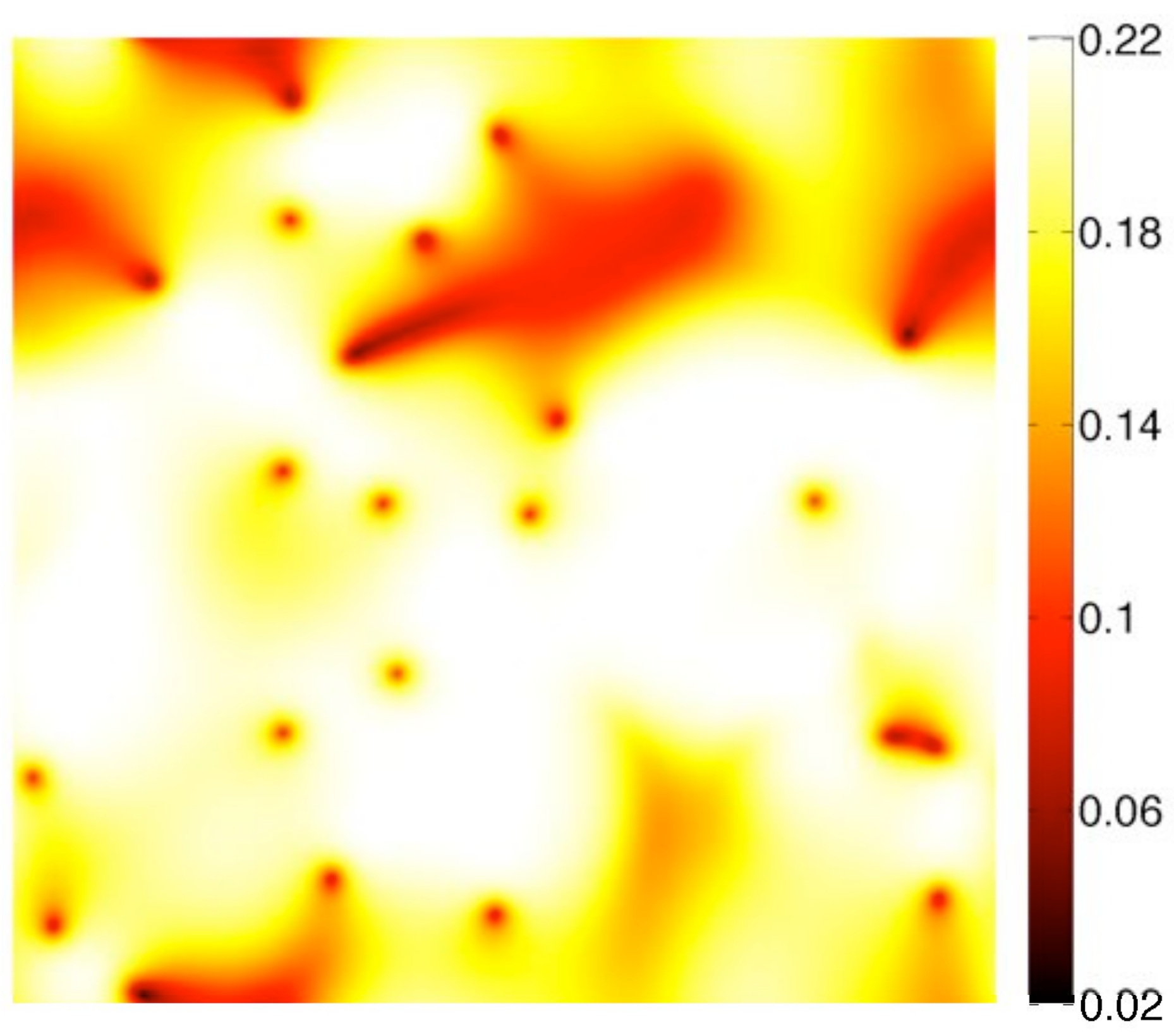}}
	\\
	\subfloat[]{\label{min}\includegraphics[scale= 0.4]{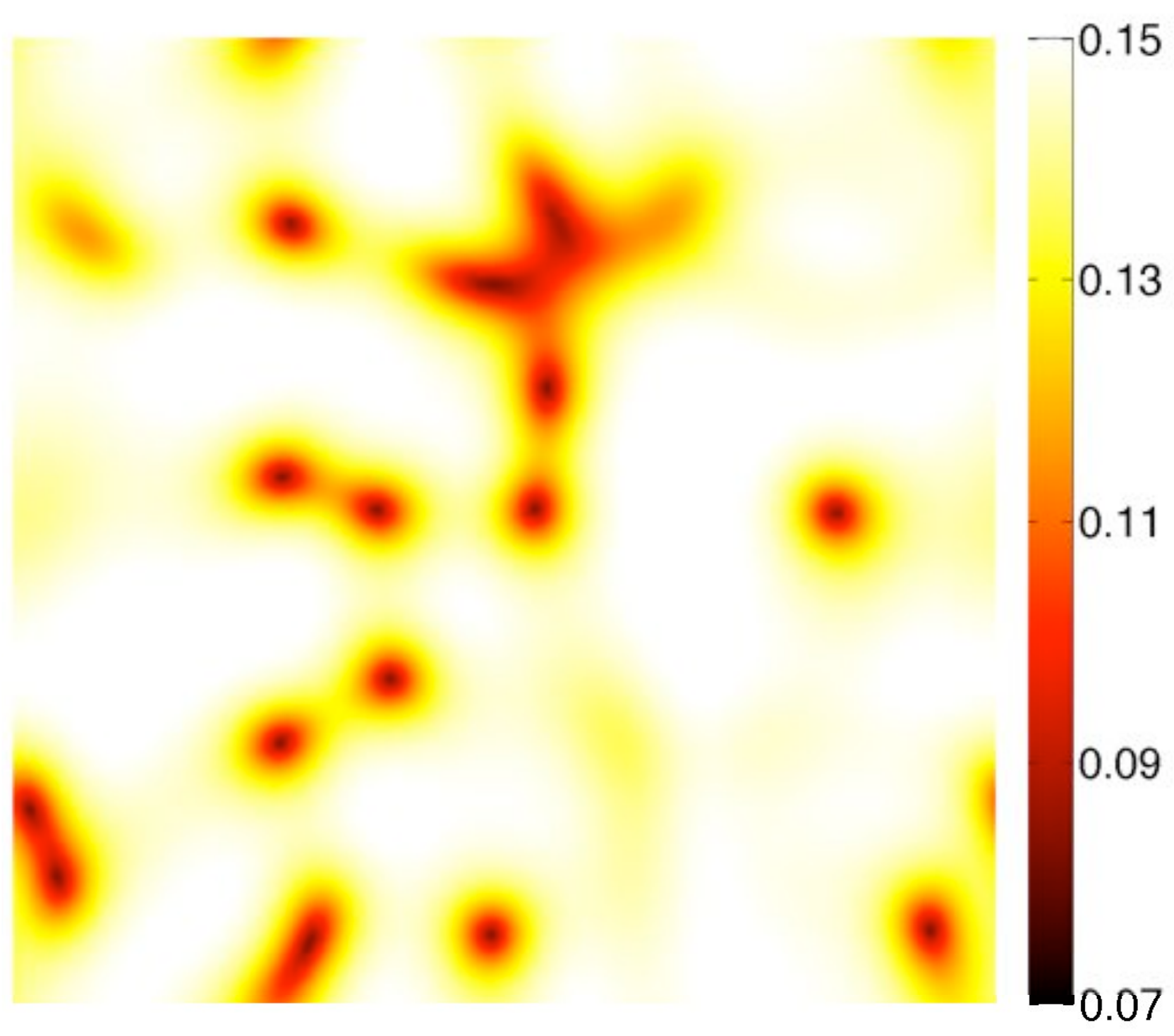}}
	\\
\caption{Figs. \ref{max} and \ref{min} represent the order parameter $S$ at every grid
point at the times corresponding to the maximum and minimum of $S$ in Fig. \ref{s_and_t_neg}, 
after the application of an electric field perpendicular to the lattice plane, for a
negative dielectric anisotropy ($\Delta \epsilon <  0$).}
  \label{neg}
\end{figure}

Figs. \ref{neg_per_cp} and \ref{neg_per_s} are analogous to Figs. \ref{before_cp} and
\ref{before_s} except that, in this case, a negative dielectric anisotropy ($\Delta \epsilon < 0$) was
considered (again with an electric field pulse perpendicular to the lattice plane). After the 
electric field is switched on the molecules (and the director ${\bf n}$)
tend to become perpendicular to the electric field, which results in the rapid suppression of integer
disclinations (only half-integer ones remain in the simulations). As expected there is a rapid
variation of the parameter $S$ (and $T$) at the half-integer charge defect cores, which can be
seen in Fig. {\ref{neg_per_s}}.

If the electric field is applied in a direction parallel to the plane of the simulation and
$\Delta \epsilon > 0$ then the director ${\bf n}$ at each grid point will become aligned
parallel to the electric field and both integer and half-integer defects are suppressed.
The most interesting situation, with the electric field parallel to the lattice plane, is that
of a negative dielectric anisotropy ($\Delta \epsilon < 0$). In this case, the director ${\bf
n}$ at each grid point becomes perpendicular to the electric field generating half-integer
twist disclinations \cite{PhysRevE.60.6831, callan-jones:061701}. The value of $\sin^{2}(2 \chi)$ is equal to
zero at every grid point but the defects can be seen in Fig. \ref{neg_par_s} which shows the
order parameter $S$ at various grid points as well as the projection of the director ${\bf n}$
onto the lattice plane. All the defects appearing in the simulations correspond to half-integer
twisted disclinations which can be identified by the values of the projections of ${\bf n}$
around them.

The evolution of the characteristic length scale, $L(t)$, is shown in Fig. {\ref{lt}}, in units of the 
simulation box size. Four different situations have been considered: $(a)$ without external 
electric field, $(b)$ with an electric field perpendicular to the lattice plane and $\Delta \epsilon > 0$, 
$(c)$ the same as $(b)$ but
with $\Delta \epsilon < 0$ and $(d)$ with an electric field parallel to the lattice plane and
$\Delta \epsilon < 0$. In each case an average over $100$ different simulations with random 
initial conditions considering has been made. The electric field is connected at the time 
$t=t_{\rm on}=400$ and disconnected at $t=t_{\rm off}=600$. In all cases the characteristic 
length scale evolves roughly as $L^{2} \propto t$ but the slope is rapidly modified when the 
electric field is switched on (or switched off in cases (c) and (d)).

Figs. \ref{pos} and \ref{s_and_t_neg} show the evolution of the parameters $S$ and $T$ for
the case of an homogeneous director, ${\bf n}$, initially aligned parallel to the lattice plane 
considering, respectively, either a positive ($\Delta \epsilon > 0$) or negative ($\Delta
\epsilon < 0$) dielectric anisotropy. In both cases an electric field perpendicular to the plane 
is connected for $t=t_{\rm on}=400$ and disconnected for $t=t_{\rm off}=600$. As expected, 
Fig. \ref{pos} shows that
the system develops a large biaxiality right after the electric field is switched on since,
initially, there are two distinct preferred directions defined by the director and the external
electric field. After some time, the director becomes aligned with the electric field ($\Delta
\epsilon >0$), $T$ is again driven towards zero and $S$ tends to larger value than before the
electric field is switched on (this is expected since the dispersion of molecular orientations
around ${\bf n}$, is reduced due to the external electric field). After the electric field is
switched off at $t=t_{\rm off}$, $S$ returns to the value it had before $t_{\rm on}$. 

For $\Delta \epsilon < 0$, Fig. \ref{s_and_t_neg} shows that again the system develops a large
biaxiality right after the electric field is generated but, in this case, the biaxiality does
not go away before $t_{\rm off}$ since the two distinct preferred directions defined by the
director and the external electric field will remain perpendicular to each other. Right after
the electric field is switched on, the value of $S$ increases sharply, before moving to a 
value smaller than before $t_{\rm on}$. The sharp growth of $S$ right after $t_{\rm on}$ is due
to an initial decrease of the dispersion of molecular orientations around ${\bf n}$ due to the
presence of the external electric field. In the subsequent evolution of $S$, before $t_{\rm off}$, 
the molecules are forced into the plane by the perpendicular electric field. This accounts for the 
fact that the value of $S(t_{\rm on})$ is slightly smaller than $S(t_{\rm off})$. Right after
$t_{\rm off}$ the value of $T$ decreases sharply towards zero, which is responsible for
corresponding sharp decrease of $S$. After that, $S$ increases again returning to the value it
had before $t_{\rm on}$.

Figs. \ref{max} and \ref{min} represent the parameter, $S$, at every grid point, at the
times corresponding to the maximum and minimum of $S$ in Fig. \ref{s_and_t_neg}, after the
application of an electric field pulse perpendicular to the lattice plane, for a negative
dielectric anisotropy ($\Delta \epsilon < 0$). They confirm the background evolution shown in
Fig. \ref{s_and_t_neg} and they also reveal a dramatic change on the characteristic scale of
defect cores which is associated with the behaviour of the characteristic scale shown in Fig.
{\ref{lt}} for cases (c) and (d).

\section{Conclusions}

In this paper we investigated the effect that an applied electric field pulse has on the dynamics of a liquid crystal 
texture network, using numerical simulations performed with LICRA,  a publicly available LIquid CRystal 
Algorithm developed by the authors. We have  demonstrated that the presence 
of an electric field can lead to significant changes of the molecular orientational distribution function, including 
the background value of the parameters $S$ ad $T$ as well as their profile at the defect cores. We have further 
shown that an electric field pulse may lead to a defect network whose properties depend on the orientation of 
the applied electric field and the sign of the dielectric anisotropy. Depending on the case, topological defects 
with integer or half integer topological charges remain after the electric pulse has been applied. In summary, our results 
describe how an applied electric field can be used to as control the nature of the defects in a liquid crystal 
texture network.

\begin{acknowledgments}

The authors would like to thank Josinaldo Menezes for useful discussions and CAPES, CNPq, REDE NANOBIOTEC BRASIL, INCT-FCx (Brasil) and FCT (Portugal) for partial support.

\end{acknowledgments}

\bibliography{cl}

\end{document}